# Parallax in the Park

Stephen W. Hughes[1], Sean Powell[1], Joshua Carroll[1], Michael Cowley[2,3]

[1]Department of Chemistry, Physics and Mechanical Engineering,
Queensland University of Technology, Gardens Point Campus,
Brisbane, Queensland 4001, Australia
[2]Department of Physics & Astronomy,
Macquarie University, Sydney, NSW 2109, Australia
[3]Australian Astronomical Observatory,
PO Box 915, North Ryde, NSW 1670, Australia

Email: sw.hughes@qut.edu.au

**Abstract**
This article describes a parallax experiment performed by undergraduate physics students at Queensland University of Technology. The experiment is analogous to the method used in astronomy to measure distances to the local stars. The result of one of these experiments is presented in this paper. A target was photographed using a digital camera at five distances between 3 and 8 meters from two vantage points spaced 0.6 m apart. The parallax distances were compared with the actual distance measured using a tape measure and the average error was $0.5 \pm 0.9$ %.

**Introduction**
Parallax is the primary method of measuring distances in the local universe. Kepler used this method to ascertain that the orbits of the planets around the Sun were elliptical, and not perfect circles, as supposed by the ancients. The main purpose of Cook's voyage to the antipodes was to observe the 1769 transit of Venus to measure the distance to Venus in km/miles using the parallax method. Up until that time the size of the Solar System was known in terms of the Astronomical Unit (AU), the average distance between the Earth and Sun, but not in km or miles. The parallax technique is relatively simple and suitable as an exercise for high school and undergraduate students (Cenadelli et al, 2009, Stewart 2011).

After the invention of photographic film, the parallax method was used to measure distances to the local stars. Ground based telescopes can measure distances within 100 pc of the earth. Space borne telescopes extend the range by a factor of 10. In 1989 the European Space Agency (ESA) launched a satellite called Hipparcos, which measured stellar distances out to 1,000 pc (sci.esa.int/hipparcos) until ceasing operations in 1993. The Hipparcos Catalogue contains accurate distances to 188,218 stars.

This paper describes a parallax experiment performed by undergraduate astrophysics students in the Brisbane city Botanical Gardens, adjacent to the Gardens Point campus of Queensland University of Technology. The technique is directly analogous to the technique used in astronomy, except photographs are taken more or less at the same time rather than six months apart. Figure 1 shows the method of measuring distances using the parallax. In the Botanic Gardens exercise, the Earth is replaced by a digital camera, the star by a bright orange target, and the fixed stars by the buildings of the Brisbane CBD.

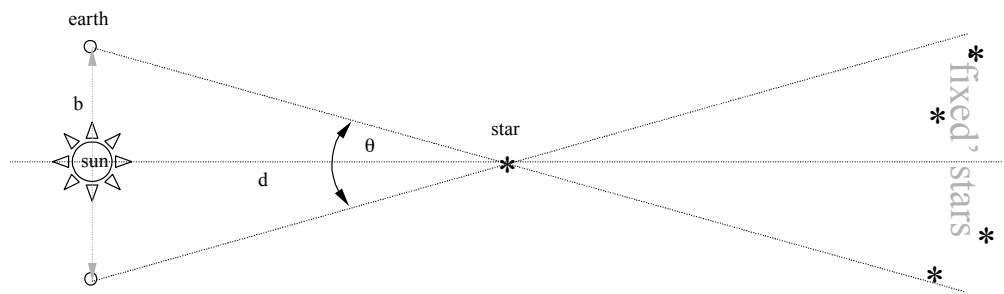

Figure 1. Measuring distances using the parallax method. The local star is photographed against the distant stars when the earth is on opposite sides of the Sun, i.e. 6 months apart. The diameter of the Earths' orbit is the baseline of a triangle with an apex angle of $\theta$. The parallax angle is $\theta/2$.

A photograph of a star is taken six months apart so that the diameter of the Earth's orbit around the Sun forms the baseline of a surveying triangle. The distance to a star is calculated using

(1)

$$d = \frac{b/2}{sin(\theta/2)}$$

Where $d$ is the distance to the star and $b$ the length of the base-line which is the diameter of the Earth's orbit around the Sun (~300 million km). The total angular shift of the star against the distant backdrop of the stars is $\theta$. The parallax angle is designated as half of the total shift, i.e. $\theta/2$.

Normally in astrophysics, the angle is specified in seconds of arc (arcseconds), which is defined as 1/3600 of a degree. The inverse of the parallax angle specified in arcseconds is the distance to a star in parsecs (pc). A star must be at a distance of 3.26 light years from the Earth to exhibit a parallax of one second of arc. The closest visible star to the Earth is Alpha Centauri at a distance of 4.4 light years (1.3 pc).

**Method**

The experiment described in this paper was performed in the Brisbane City Botanic Gardens adjacent to the Gardens Point campus of Queensland University of Technology. The photographs were taken by a group of second year physics students, although author SH analysed the images independently and obtained similar results.

Two tripods were placed on the lawn. The distance between the camera attachments was 60 ± 1 cm. Initially, a meter rule was placed on a tripod 3.38 m from the camera. The photo is shown in figure 2. This image was used to calculate the physical size of the pixels in the CCD chip in the camera.

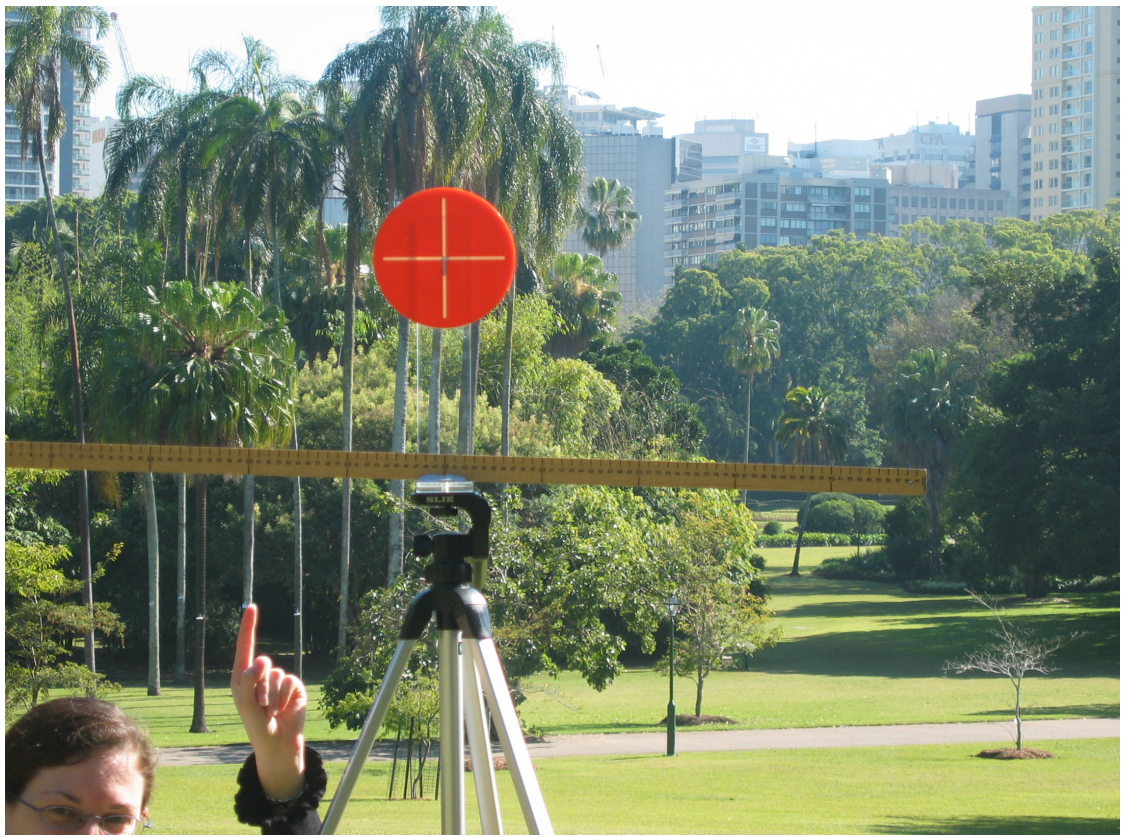

Figure 2. Photo of a metre rule on a tripod for obtaining the physical size of the CCD pixels in the camera.

The target shown in figure 3 was placed at a distance 3.38, 4.78, 5.69, 6.82 and 7.82 ± 0.01 m from the midpoint of the line between the tripod camera attachment bolts. A photo of the target was taken with a Canon Power Shot S45 4 Mega Pixel camera placed first on one tripod and then on the other. The focus was set to infinity so that both the target and buildings of the CBD were in focus. The camera does not have to point exactly at the target.

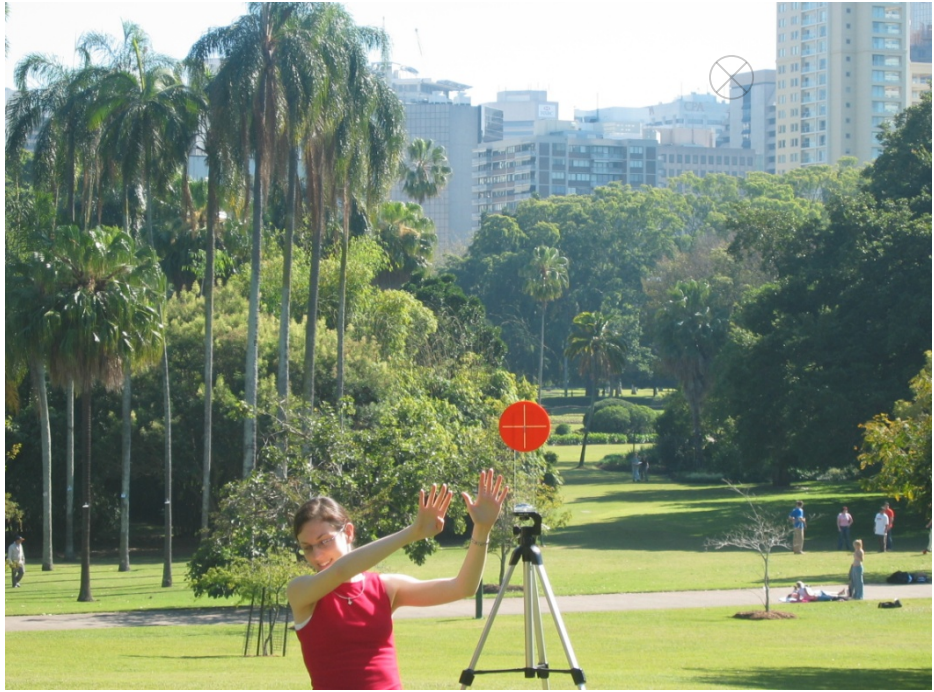

Figure 2. A QUT student performing the parallax experiment. The display of nine fingers indicates that the photo is the 9$^{th}$ image in the sequence. In this case the target was 7.8 m away from the camera baseline. The diagonal cross in the circle in the top right of the picture is the origin used for calculating the parallax.

After the experiment the photos were placed on the university Blackboard teaching website to enable students to download the images for further analysis. The first step is to calculate the physical size of the CCD pixels is calculated as shown in figure 3 and following equations.

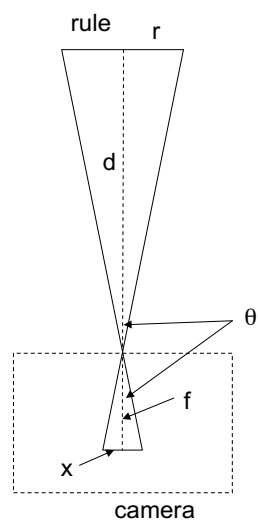

Figure 3. Schematic diagram of the experimental geometry.

$$\frac{x}{f} = \frac{r}{d} \Rightarrow x = \frac{rf}{d} \qquad \text{pixel size} = \frac{x}{n} \qquad \theta = \tan^{-1}\left(\frac{x}{f}\right)$$

Where $f$ = focal length of the lens, $x$ = physical distance of a point on the image from the centre-line, $d$ = distance along centre-line to rule, $r$ = distance from centre-line to a point on the rule, $n$ = no of pixels from centre-line.

The one metre rule was photographed at a distance of 3.38 ± 0.01 m with a camera focal length of 21.3225 mm. The horizontal distance of a 50 cm section of the rule was 980 pixels. Therefore, the physical pixel size on the CCD chip was

$$x = \frac{rf}{d} = \frac{500 \times 21.3225}{3380} = 3.154 \pm 0.009 \, mm$$

$$pixel\ size = \frac{x}{n} = \frac{3.154}{980} = 0.003218 \pm 0.000016 \, mm = 3.218 \pm 0.016 \, \mu m$$

The error in $r$ was taken to be ± 1 cm and the error in $n$ ± 2 pixels. The error in the pixel size was calculated by reducing the distance to the rule from 3.38 to 3.37 m and the horizontal pixel length from 980 to 978. The length of the rule used in the calibration was assumed to be fixed at 0.5 m and the camera focal length 21.3225 mm as specified in the image header.

For each pair of images (2-3, 4-5, 6-7, 8-9,10-11) the following procedure was used. A reference point, or origin, was chosen in the image – as marked on the image in figure 2. The x coordinate of the origin and centre of the target was recorded and the target subtracted from the origin (Δx). This was repeated for the second image. (N.B. the ordering of each pair of images is not important). The two Δx values were then subtracted from each other and the modulus taken Δ(Δx), as shown in table 1.

For each Δ(Δx), the physical CCD distance was found by multiplying by the pixel size. This was used in conjunction with the lens focal length to calculate the parallax angle using equation (1). When the errors in the measurements were propagated the overall error was found to be ± 17 cm. This is the size of the error bars on the plot in figure 5. Note that the linear regression line passes through each error bar indicating that this is a reasonable estimate of the error.

The group of students who acquired the images shown in this paper used a slightly different method to calculate the parallax distance. They matched the distant background in each pair of images and calculated the horizontal pixel distance directly as shown in figure 4.

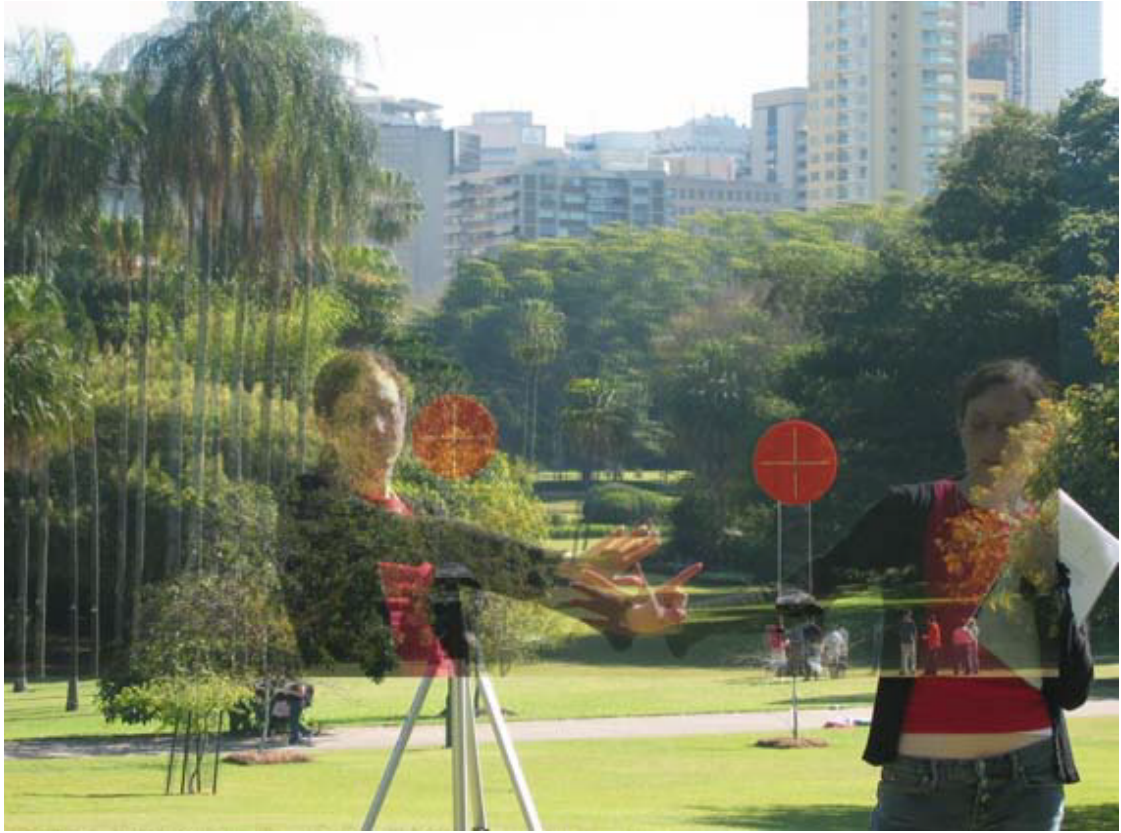

Figure 4. Two photos have been overlayed by matching the distant background to enable direct measurement of the horizontal distance between the target in the two images.

The pixel distance was converted to a physical distance by multiplying the pixel distance by the physical size of each CCD element. The parallax angle was then calculated using the focal length of the lens in the equation $\theta = tan^{-1}(\Delta x/f)$. The parallax distance was then calculated using equation (1).

**Results**

Table 1 shows the data obtained from five pairs of images (provided in supplementary materials). Table 2 shows the measurements in a more convenient form with the associated percentage error for each pair of values. Figure 5 shows a plot of parallax distance versus actual distance with associated error bars.

Table 1. The basic data obtained from the images.

| Tape (m) |        | Image 1 | Image 2 | Δ(Δx) | x (mm) | θ (rads) | Parallax (m) |
|---:|---|---:|---:|---:|---:|---:|---:|
| 7.82 | Target | 1272 | 1766 | | | | |
|  | Origin | 1778 | 1760 | | | | |
|  | Δx | -506 | 6 | 512 | 1.65 | 0.077 | 7.76 |
|  |  |  |  |  |  |  |  |
| 6.82 | Target | 1668 | 1218 | | | | |
|  | Origin | 1646 | 1778 | | | | |
|  | Δx | 22 | -560 | 582 | 1.87 | 0.088 | 6.83 |
|  |  |  |  |  |  |  |  |
| 5.69 | Target | 916 | 1732 | | | | |
|  | Origin | 1524 | 1646 | | | | |
|  | Δx | -608 | 86 | 694 | 2.23 | 0.105 | 5.73 |
|  |  |  |  |  |  |  |  |
| 4.78 | Target | 1950 | 826 | | | | |
|  | Origin | 1818 | 1536 | | | | |
|  | Δx | 132 | -710 | 842 | 2.71 | 0.127 | 4.72 |
|  |  |  |  |  |  |  |  |
| 3.38 | Target | 894 | 1939 | | | | |
|  | Origin | 1976 | 1828 | | | | |
|  | Δx | -1082 | 111 | 1193 | 3.84 | 0.180 | 3.33 |

Table 2. Percentage error of parallax distance measurements

| Tape (m) | Parallax (m) | % error |
|---:|---:|---:|
| 3.38 | 3.33 | -1.34 |
| 4.78 | 4.72 | -1.22 |
| 5.69 | 5.73 | 0.66 |
| 6.82 | 6.83 | 0.13 |
| 7.82 | 7.76 | -0.75 |
|  | mean | 0.50 |
|  | sd | 0.87 |

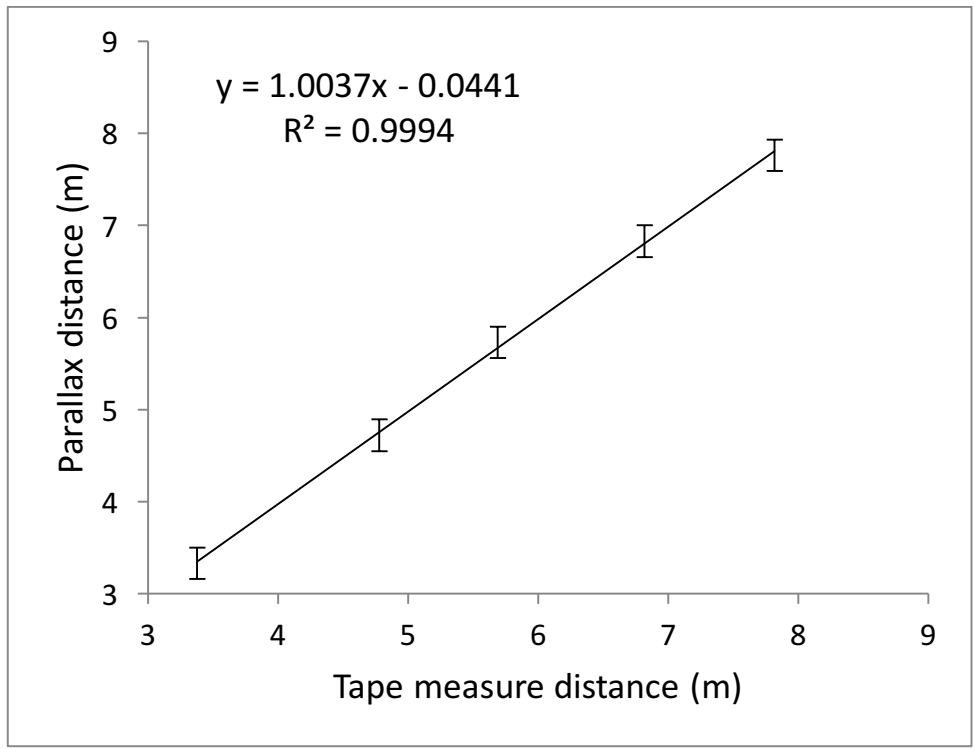

Figure 5. Plot of parallax distance verses the distance measured using a tape measure. The error bars are ± 17 cm.

**Discussion**

This parallax experiment can achieve surprisingly accurate results. No doubt the results could be improved, but this is probably not necessary since the purpose of this experiment is to give students hands-on experience of measuring distance using parallax.

**References**

Cenadelli D, Bernagozzi A, Calcidese P, Faerreira L, Hoang C, Rijsdijk C An international parallax campaign to measure distance to the Moon and Mars *Eur. J. Phys.* **30** 35-46.

Stewart B 2011 Measuring stellar distances by parallax *Phys. Educ.* **46** 137 – 138.